\documentclass[final,3p,times,sort&compress]{elsarticle}

\usepackage{multirow,setspace,times,amssymb,amsmath,graphicx,color,rotating,subfigure,url,epstopdf}
\usepackage{lineno}
\usepackage{natbib}
\usepackage{array}
\usepackage{diagbox}
\usepackage[colorlinks,linkcolor=red,anchorcolor=blue,citecolor=green]{hyperref}
\usepackage{bigstrut}

\usepackage{dcolumn}
\newcolumntype{d}[1]{D{.}{.}{#1}}

\journal{Elsevier}

\begin{document}

\begin{frontmatter}

\title{{Visibility graph analysis of economy policy uncertainty indices}}

\author[CME]{Peng-Fei Dai}
\author[CME,CCSCA]{Xiong Xiong}
\author[BS,RCE,SS]{Wei-Xing Zhou \corref{cor1}}
\cortext[cor1]{Corresponding author.}
\ead{wxzhou@ecust.edu.cn} %

\address[CME]{College of Management and Economics, Tianjin University, Tianjin 300072, China}
\address[CCSCA]{China Center for Social Computing and Analytics, Tianjin University, Tianjin 300072, China}
\address[BS]{Department of Finance, East China University of Science and Technology, Shanghai 200237, China}
\address[RCE]{Research Center for Econophysics, East China University of Science and Technology, Shanghai 200237, China}
\address[SS]{Department of Mathematics, East China University of Science and Technology, Shanghai 200237, China}

\begin{abstract}
Uncertainty plays an important role in the global economy. In this paper, the economic policy uncertainty (EPU) indices of the United States and China are selected as the proxy variable corresponding to the uncertainty of national economic policy. By adopting the visibility graph algorithm, the four economic policy uncertainty indices of the United States and China are mapped into complex networks, and the topological properties of the corresponding networks are studied. The Hurst exponents of all the four indices are within $\left[0.5,1\right]$, which implies that the economic policy uncertainty is persistent. The degree distributions of the EPU networks have power-law tails and are thus scale-free. The average clustering coefficients of the four EPU networks are high and close to each other, while these networks exhibit weak assortative mixing. We also find that the EPU network in United States based on daily data shows the small-world feature since the average shortest path length increases logarithmically with the network size such that $L\left(N\right)=0.626\ln N+0.405$. Our research highlights the possibility to study the EPU from the view angle of complex networks.
\end{abstract}

\begin{keyword}
Econophysics; Economic policy uncertainty; Complex network; Visibility graph
\PACS 89.65.Gh \sep 89.75.Da \sep 02.50.-r \sep 89.90.+n \sep 05.65.+b
\end{keyword}


\end{frontmatter}


\section{Introduction}

Nowadays, the development of world economy has been highly globalized, and the ties between different economies are getting stronger and stronger. The internal factors and external environment that affect economic development are changing over time. Hence, uncertainty has become a new normal that the global economy needs to face. Particularly, economic policy uncertainty (EPU) is related to the development of financial markets and even the harmony and stability of the society. The research on economic policy uncertainty becomes a hot topic in the fields of macro-economy and macro-finance.

The study on economic policy uncertainty mainly focuses on theoretical research and econometric analysis. P{\'{a}}stor and Veronesi developed a general equilibrium model to analyze how changes in government policy choice affect stock prices \cite{Veronesi-Pastor-2012-JF}. In their model, two types of uncertainty are considered, the political uncertainty and the impact uncertainty. P{\'{a}}stor and Veronesi further set up a general equilibrium model of government policy choice to study the relationship between political uncertainty and stock risk premium \cite{Veronesi-Pastor-2013-JFE}. In their study, some simple exploratory empirical analyses were conducted, where the economic policy uncertainty index constructed by Baker et al. was adopted as the proxy variable of policy uncertainty \cite{Baker-Bloom-Davis-2016-QJE}. Segal et al. decomposed the aggregate uncertainty into `good' and `bad' volatility components, associated with positive and negative innovations to macroeconomic growth \cite{Segal-Shaliastovich-Yaron-2015-JFE}. According to their theoretical analysis, the two kinds of uncertainty have opposite effects on economic growth and asset prices. Baker et al. developed a new index of economic policy uncertainty, and made empirical analyses of the impact of economic policy uncertainty on micro-finance and macro-economy from the perspectives of firm levels and macro levels \cite{Baker-Bloom-Davis-2016-QJE}. Jes\'us et al. studied empirically the effects of changes in uncertainty about future fiscal policy on the aggregate economic activity \cite{Jesus-Pablo-Keith-Juan-2015-AER}. Brogaard and Detzel used a search-based measure to capture economic policy uncertainty for 21 economies, and found economic policy uncertainty has a significant effect on the contemporaneous market returns and volatility \cite{Brogaard-Detzel-2015-MS}. Dakhlaoui and Aloui found strong evidence of a time-varying correlation between US economic uncertainty and stock market volatility \cite{Dakhlaoui-Aloui-2016-IE}. What is worth mentioning is that the studies of the EPU index attract the interest of a large number of scholars since Baker et al. introduced the concept \cite{Baker-Bloom-Davis-2016-QJE}.


Alternatively, it is possible to study the EPU time series from the view angle of complex networks. Indeed, quite a few methods have been developed to convert time series into complex networks \cite{Zou-Donner-Marwan-Donges-Kurths-2019-PR,Gao-Li-Dang-Yang-Do-Grebogi-2017-IJBC,Gao-Zhang-Dang-Yang-Wang-Duan-Chen-2018-KBS,Gao-Dang-Mu-Yang-Li-Grebogi-2018-IEEEtii}. Zhang and Small connected time series with complex networks, by constructing complex networks from pseudoperiodic time series, and investigate the relationship between the topology of the constructed network and dynamics of the raw time series \cite{Zhang-Small-2006-PRL,Zhang-Sun-Luo-Zhang-Nakamura-Small-2008-PD}. Yang et al. and Gao et al proceeded to construct segment correlation networks \cite{Yang-Yang-2008-PA,Gao-Jin-2009-PRE}. Xu et al. designed the method to map time series into nearest neighbor networks \cite{Xu-Zhang-Small-2008-PNAS}. Marvan et al. introduced the concept of recurrence networks \cite{Marwan-Donges-Zou-Donner-Kurths-2009-PLA,Donner-Zou-Donges-Marwan-Kurths-2010-NJP}.
Lucasa et al. introduced the visibility graph (VG) algorithm for time series \cite{Lacasa-Luque-Ballesteros-Luque-Nuno-2008-PNAS}, and various expansions of the visibility graph algorithm have been proposed \cite{Luque-Lacasa-Ballesteros-Luque-2009-PRE,Lacasa-Luque-Luque-Nuno-2009-EPL,Ahadpour-Sadra-2012-IS,Zhou-Jin-Gao-Luo-2012-APS,Ahadpour-Sadra-ArastehFard-2014-IS,Chen-Hu-Mahadevan-Deng-2014-PA,Bezsudnov-Snarskii-2014-PA,Zou-Donner-Marwan-Small-Kurths-2014-NPG,Gao-Yang-Fang-Zou-Xia-Du-2015-EPL,Lacasa-Nicosia-Latora-2015-SR,Gao-Cai-Yang-Dang-Zhang-2016-SR,Snarskii-Bezsudnov-2016-PRE,Gao-Cai-Yang-Dang-2017-PA,Bianchi-Livi-Alippi-Jenssen-2017-SR,Xu-Zhang-Deng-2018-CSF}. At the same time, in-depth analyses of these VG-based methods and their applications appeared in different fields \cite{Ni-Jiang-Zhou-2009-PLA,Yang-Wang-Yang-Mang-2009-PA,Liu-Zhou-Yuan-2010-PA,Qian-Jiang-Zhou-2010-JPA,Xie-Zhou-2011-PA,Wang-Li-Wang-2012-PA,Yu-2013-PA,Ravetti-Carpi-Goncalves-Frery-Rosso-2014-PLoS1,Sun-Wang-Gao-2016-PA,Xie-Han-Jiang-Wei-Zhou-2017-EPL,Zhang-Zou-Zhou-Gao-Guan-2017-CNSNS,Zhang-Shang-Xiong-Xia-2018-FNL,Vamvakaris-Pantelous-Zuev-2018-PA,Xie-Han-Zhou-2019-EPL,Nguyen-Nguyen-Nguyen-2019-PA,Fan-Li-Yin-Tian-Liang-2019-AEn,Wang-Zheng-Wang-2019-PA,Xie-Han-Zhou-2019-CNSNS}.


In this paper, we convert the time series of EPU indices into complex networks by adopting the visibility graph algorithm and study the topological properties of the constructed complex networks. The remainder of the paper is organized as follows. Section~\ref{S1:Data} describes the data we analyze. Section~\ref{S1:Method} presents briefly the visibility graph algorithm. In Section~\ref{S1:EmpAnal}, the topological properties of the uncertainty networks are studied in detail. Section~\ref{S1:Conclude} concludes.

\section{Data description}
\label{S1:Data}

The EPU index is the proxy variable of economic policy uncertainty. Baker et al. constructed the index and update the data regularly on a website (\url{http://www.policyuncertainty.com/}) \cite{Baker-Bloom-Davis-2016-QJE}. To measure policy-related economic uncertainty of the United States, they constructed an index from three types of underlying components. The first component quantifies the newspaper coverage of policy-related economic uncertainty. The second component reflects the number of federal tax code provisions set to expire in future years. The third component uses disagreement among economic forecasters as a proxy for uncertainty. To measure the EPU for China, they constructed a scaled frequency count of articles about policy-related economic uncertainty in the South China Morning Post~(SCMP), one of Hong Kong's leading English-language newspapers. The method follows the news-based EPU index for the United States.

\begin{figure}[htbp]
\centering
\includegraphics[width=0.45\linewidth]{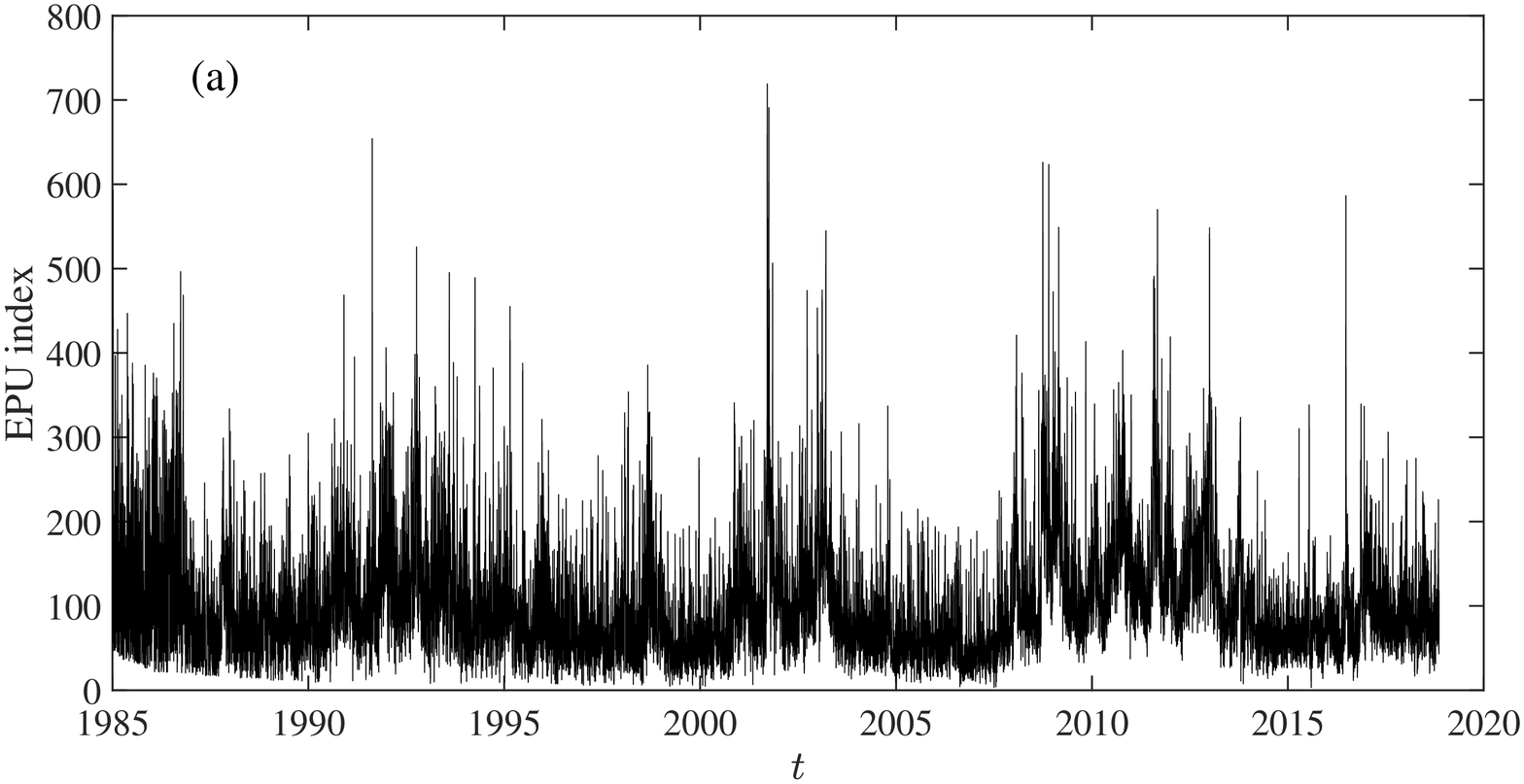}
\includegraphics[width=0.45\linewidth]{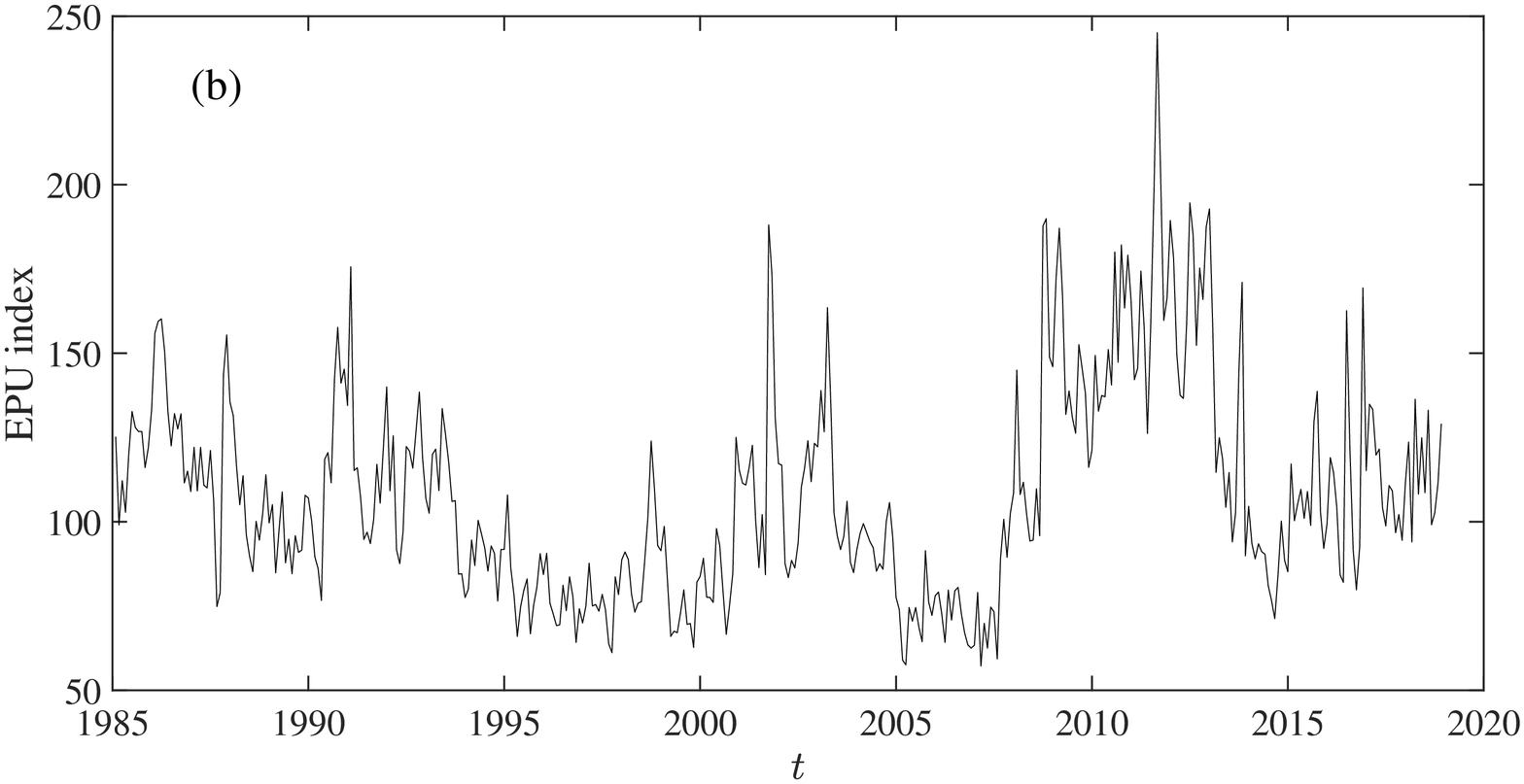}
\includegraphics[width=0.45\linewidth]{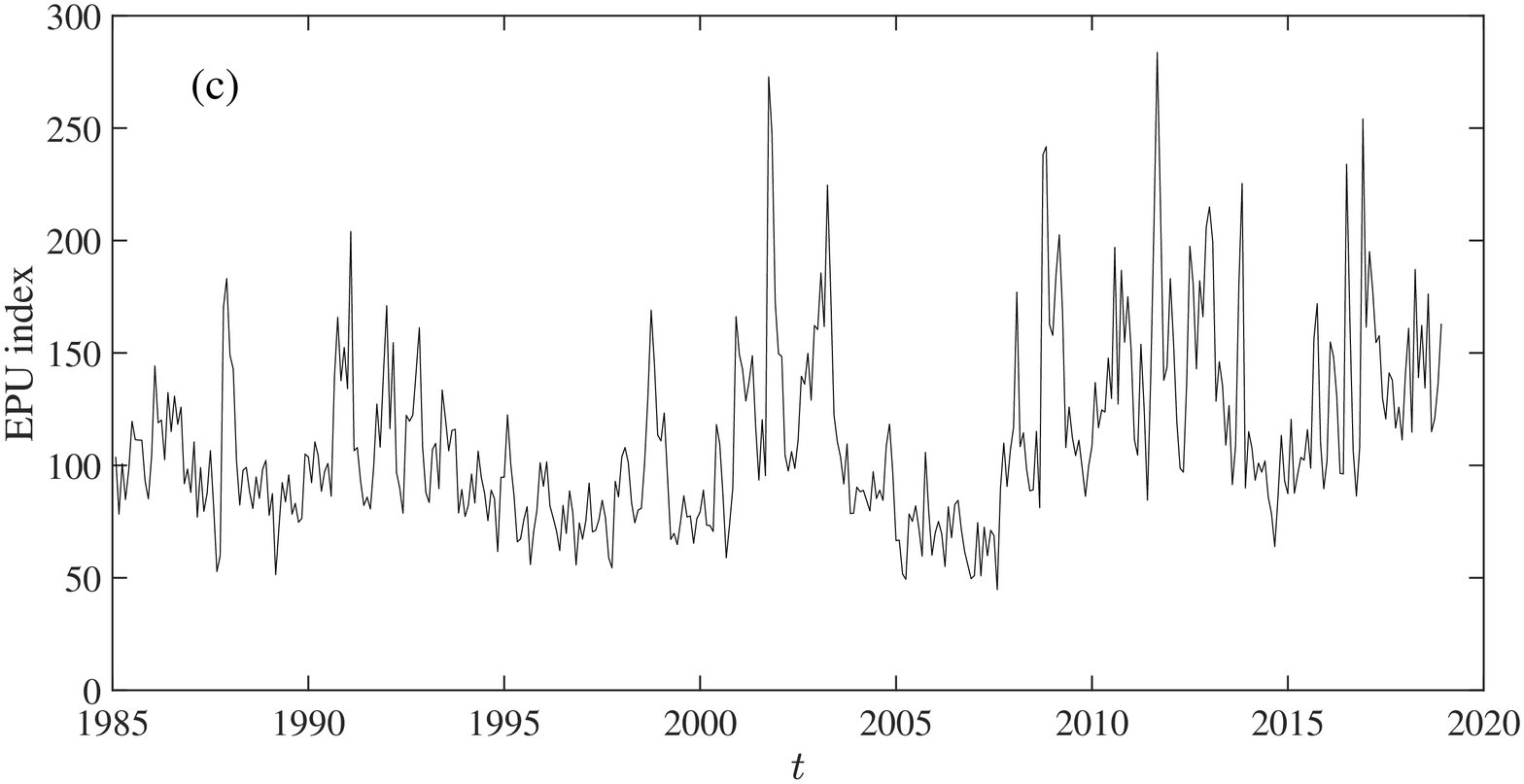}
\includegraphics[width=0.45\linewidth]{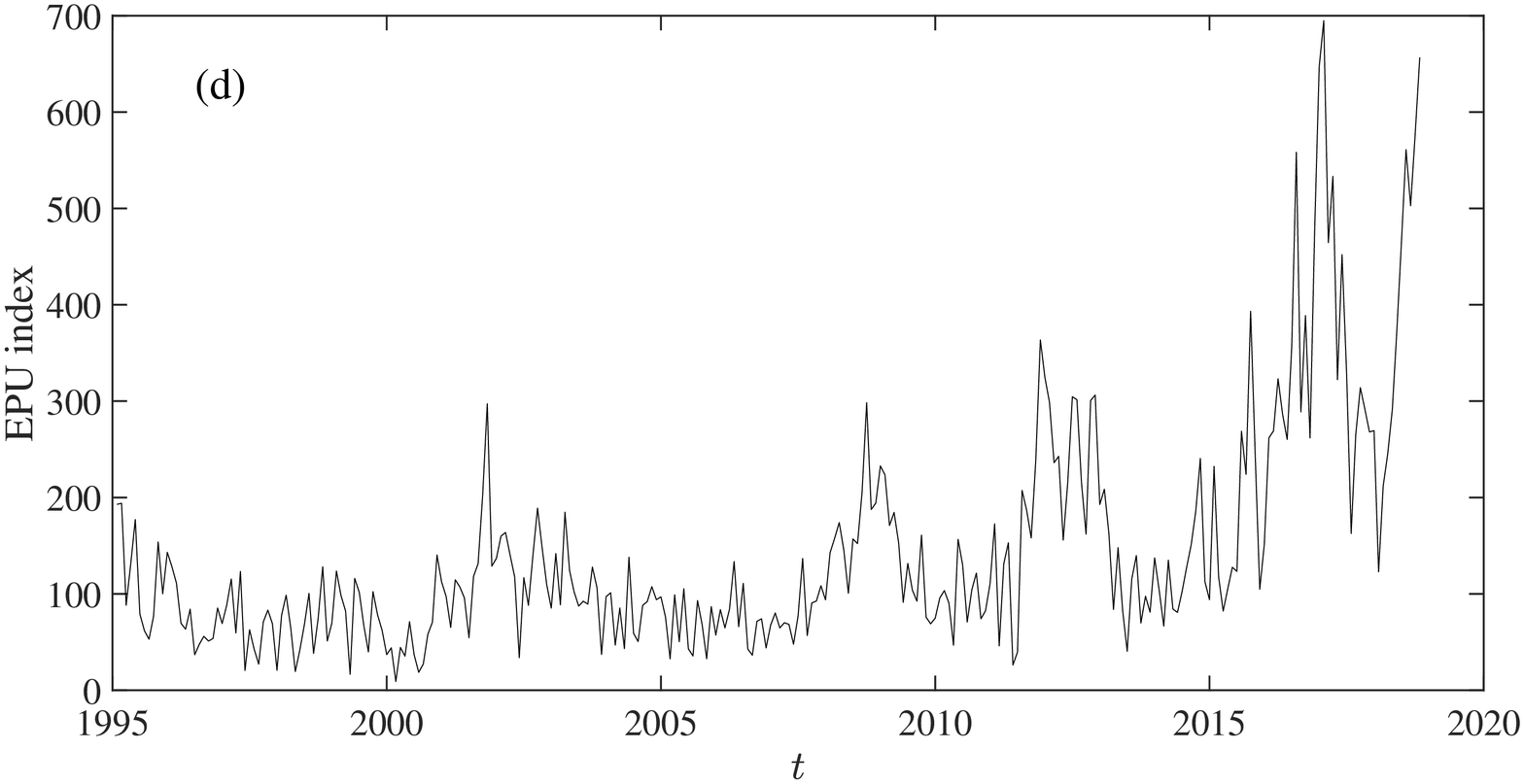}
\caption{Temporal evolution of the four EPU indices: (a) daily EPU index of the USA (EPU-US-D), (b) monthly EPU index of the USA (EPU-US-M), (c) news-based monthly EPU index of the USA (EPU-US-News-M), and (d) monthly EPU index of China (EPU-CN-M).}
\label{Fig:EPUNetStat:RawData}
\end{figure}

We retrieved the EPU indices of the United States and China, the world's two largest economies. The EPU indices of the USA include monthly data, from January 1985 to November 2018, and daily data, from 1 January 1985 to 11 November 2018. The monthly data comprise the EPU index and the news-based EPU index. In contrast, the daily data are only news-based EPU index. The EPU index of China contains only the monthly data, and its range is from January 1985 to November 2018. Figure~\ref{Fig:EPUNetStat:RawData} illustrates the temporal evolution of the four EPU indices.  

Table~\ref{TB:EPUNetStat:DataStat} shows the summary statistics of the EPU indices, including their mean, median, minimum, maximum, standard deviation, skewness and kurtosis. According to Table~\ref{TB:EPUNetStat:DataStat}, all the three mean values of the US EPU indices are close to the benchmark value of 100, while that of China is 43.75\% above the benchmark value. By comparing the standard deviations of the monthly EPU indices in the two countries, we can find that the volatility of China's EPU inde is much higher than that of the United States. The monthly EPU index of the United States (EPU-US-M) ranges from 57.21 to 245.13, while the news-based monthly US EPU index (EPU-US-News-M) varies from 44.78 to 283.67. China's monthly EPU index (EPU-CN-M) varies widely, from 9.067 to 694.849. Furthermore, the data in the United States cover a larger time range than that in China. This shows that from January 1995 to November 2018, the variation of EPU index in the United States is more stable than that in China and the economic policy uncertainty of China is higher and more volatile than that of the United States. Analysing only the statistical results of the United States, we find that the news-based EPU-US-News-M is almost the same as the EPU index in terms of mean, standard deviation, skewness and kurtosis. 

\begin{table}[htp]
\centering
\caption{Summary statistics of the EPU indices.}
\smallskip
\label{TB:EPUNetStat:DataStat}
\begin{tabular}{cd{5.0}cd{3.2}d{3.2}cd{3.2}cc}
  \hline \hline
 Variable      &   \multicolumn{1}{c}{Observation} &   Mean  &  \multicolumn{1}{c}{Median}  &   \multicolumn{1}{c}{Minimum} & Maximum  &     \multicolumn{1}{c}{Std. dev.}  & Skewness & Kurtosis \\ \hline
 EPU-US-D      &  12368 &  100.93  &  83.63  &   3.32  &  719.07 &   68.43  & 1.86    & 5.95 \\
 EPU-US-M      &  407   &  108.11  & 102.20  &  57.20  &  245.13 &   31.30  & 0.96    & 0.86 \\
 EPU-US-News-M &  407   &  111.26  & 102.02  &  44.78  &  283.67 &   40.10  & 1.27    & 1.95 \\
 EPU-CN-M      &  286   &  143.75  & 104.43  &   9.07  &  694.85 &  117.16  & 2.22    & 5.32 \\ \hline
\end{tabular}
\end{table}


%
%
%

\section{The visibility algorithm}
\label{S1:Method}

The visibility graph algorithm maps a time series $\{y_i=y(t_i)\}_{i=1,\ldots,N}$ into a visibility graph $G$ \cite{Lacasa-Luque-Ballesteros-Luque-Nuno-2008-PNAS}, where each data value $(y_i)$ in the time series corresponds to a data point $(t_i,y_i)$, which is viewed as a node in the visibility graph. Any two nodes in the visibility graph are connected if they are visible to each other. Mathematically, two points $(t_i,y_i)$ and $(t_j,y_j)$ are visible to each other if and only if:
\begin{equation}
\frac{y_j-y_n}{t_j-t_n} > \frac{y_j-y_i}{t_j-t_i},
\end{equation}
for any $t_n\in(t_i,t_j)$.

According to the construction manner, each point is at least connected to its two neighbor points. Therefore every visibility graph is connected. In this basic setting, the edges in the network are not weighted, and there are no directions for the edges. That is, the constructed networks are unweighted and undirected.

%
%

\section{Empirical analysis}
\label{S1:EmpAnal}

\subsection{Hurst exponents of the EPU time series}

\begin{figure}[htbp]
\centering
\includegraphics[width=0.45\linewidth]{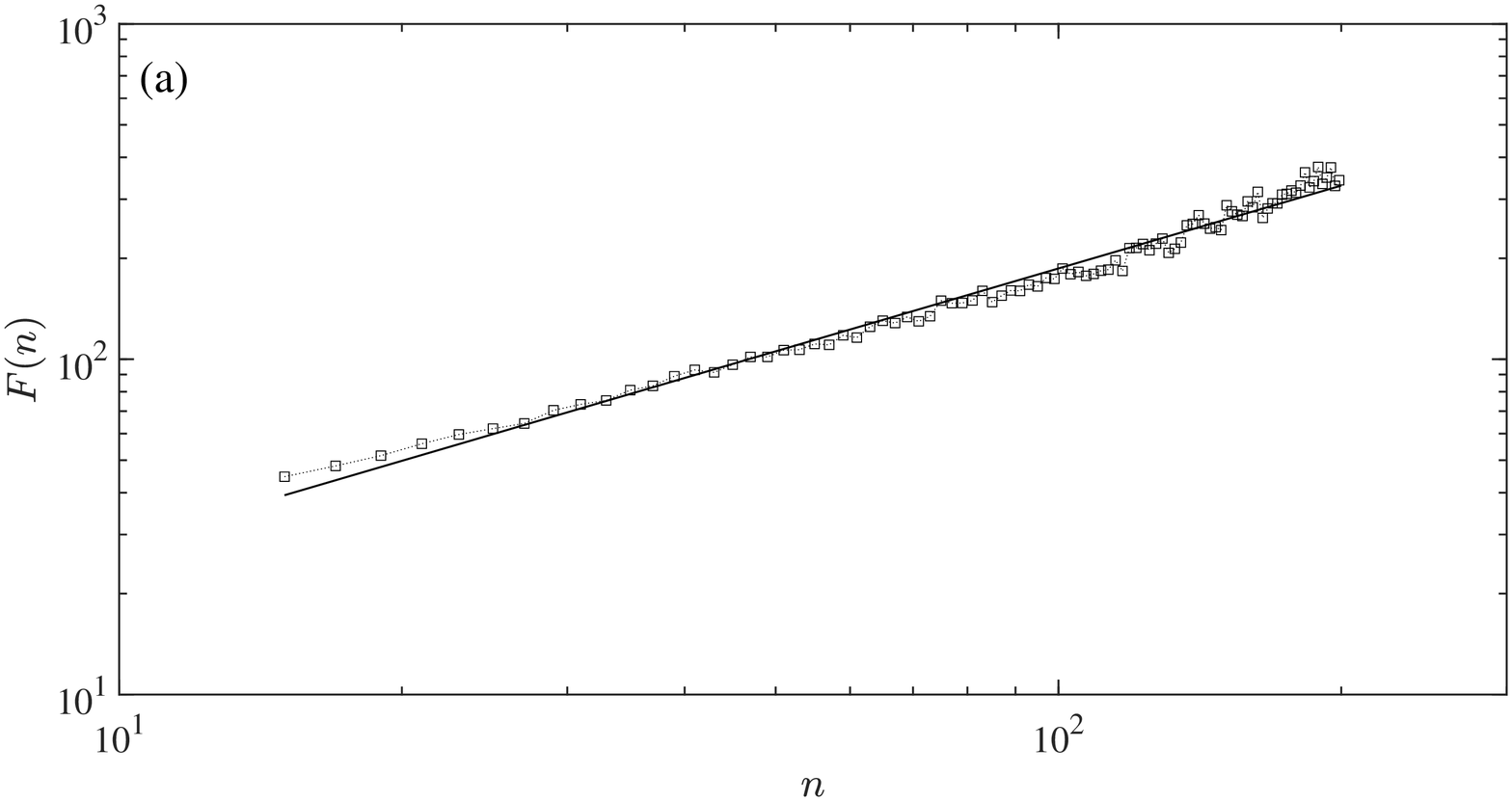}
\includegraphics[width=0.45\linewidth]{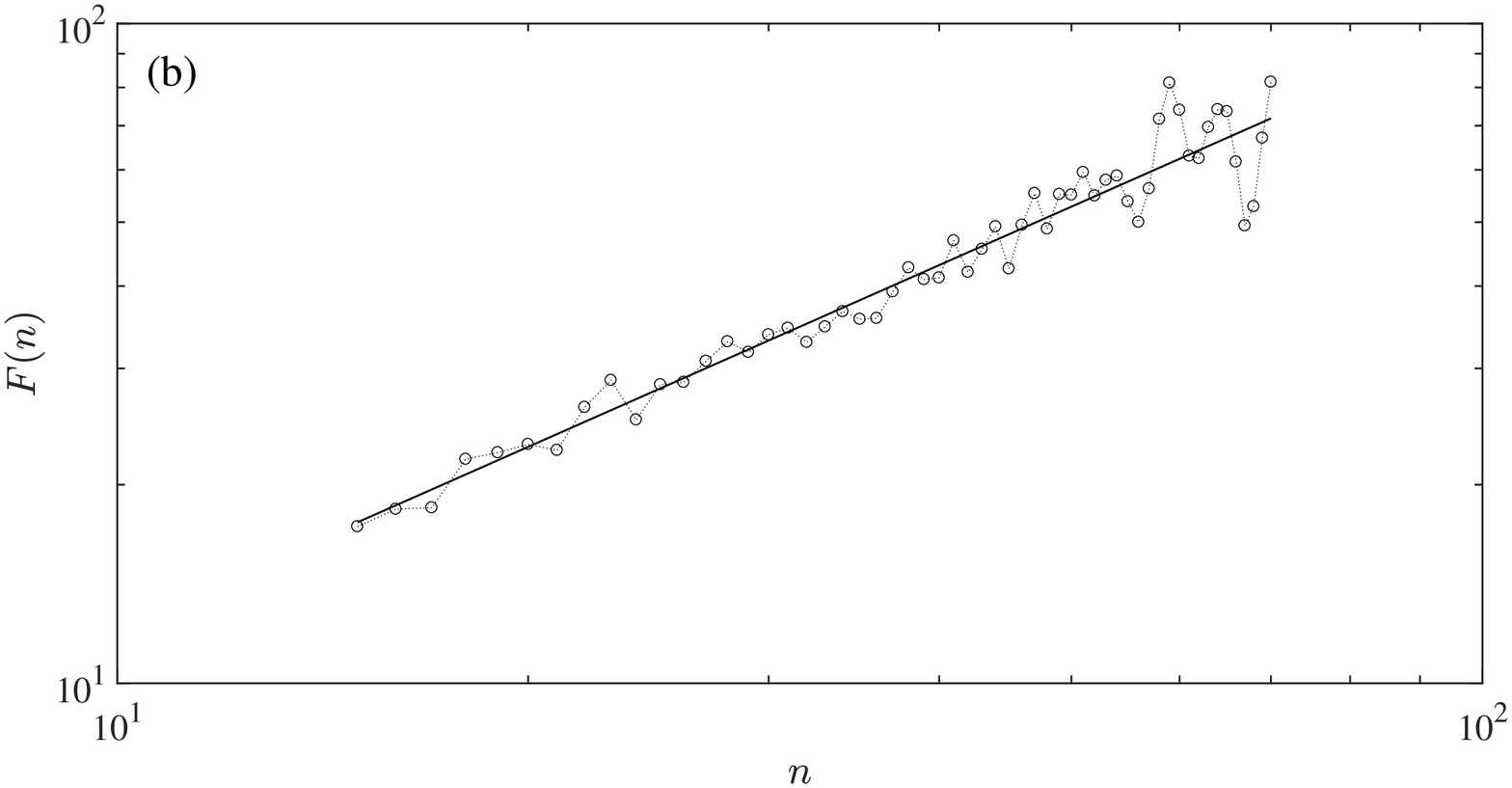}
\includegraphics[width=0.45\linewidth]{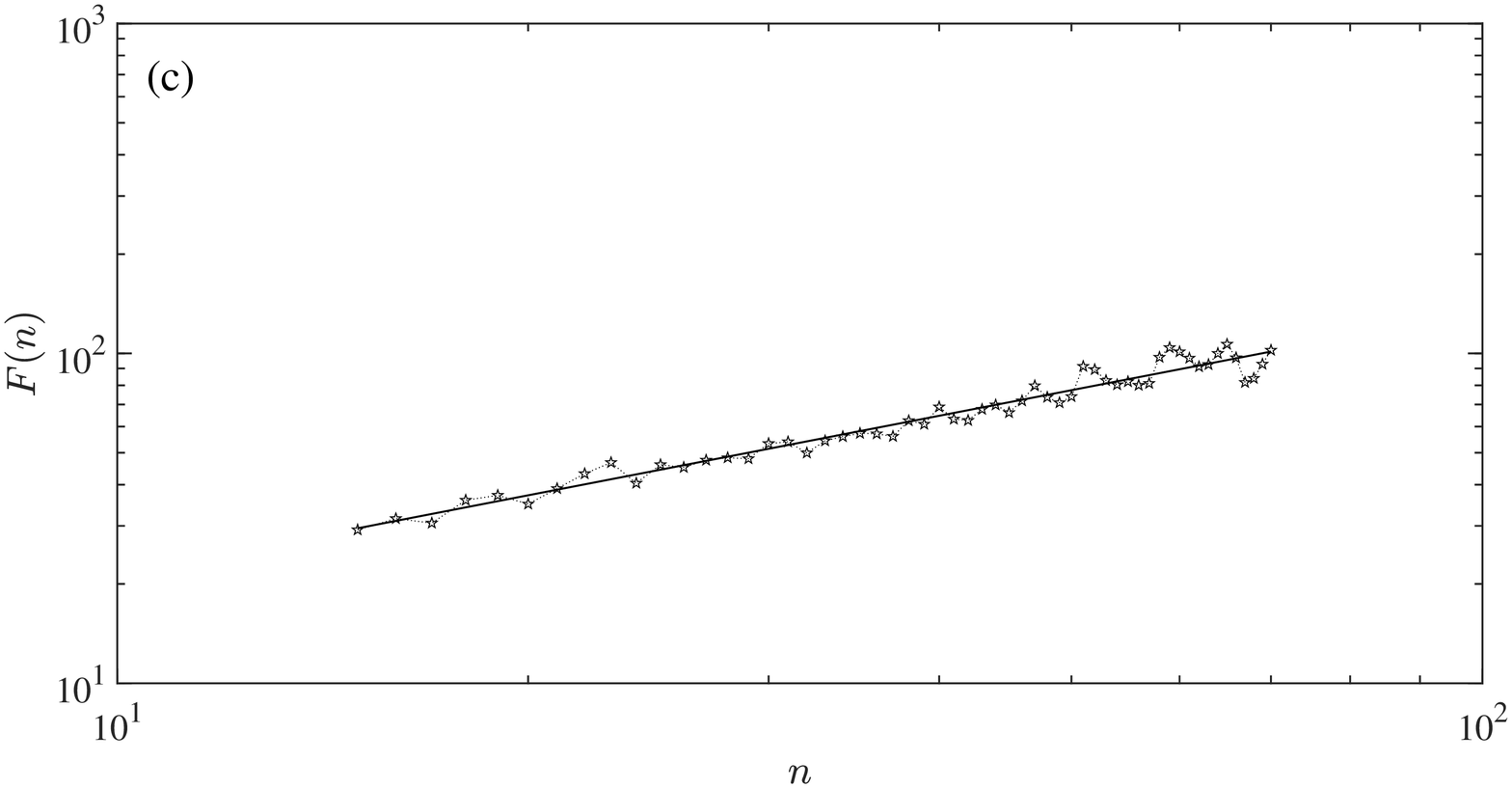}
\includegraphics[width=0.45\linewidth]{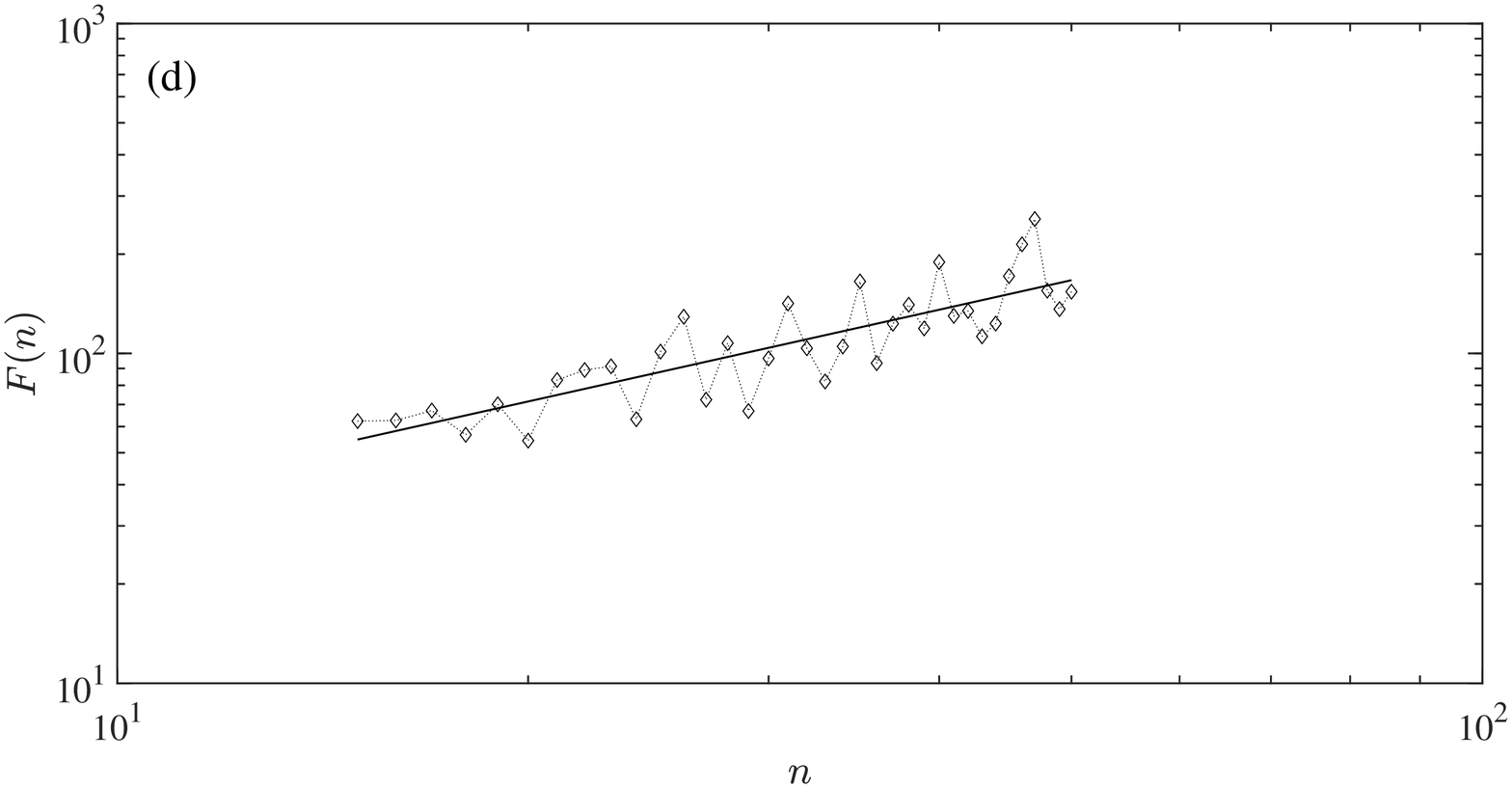}
\caption{The scaling behavior curves of the four EPU indices: (a) EPU-US-D, (b) EPU-US-M, (c) EPU-US-News-M, and (d) EPU-CN-M, where $n$ is the timescale and $F(n)$ is the fluctuation function.}
\label{Fig:EPUNetStat:Hurst}
\end{figure}

The Hurst exponent can be used to measure whether a time series is long-range correlated or not and is also an important indicator to describe the fractal characteristics of the time series. Time series can be classified into three categories due to their auto-correlation features quantified by their Hurst exponents: A time series is anti-persistent if $0<H<0.5$, uncorrelated if $H=0.5$, or persistent if $0.5<H<1$. We adopt the derended fluctuation analysis (DFA) to estimate the Hurst exponent of each EPU index, where quadratic polynomial is employed \cite{Peng-Buldyrev-Havlin-Simons-Stanley-Goldberger-1994-PRE,Kantelhardt-KoscielnyBunde-Rego-Havlin-Bunde-2001-PA,Jiang-Xie-Zhou-Sornette-2018-XXX}. For the sake of illustration, the scaling behavior curves of the four EPU indices are presented in Fig.~\ref{Fig:EPUNetStat:Hurst}.

The Hurst exponent is 0.835 for the EPU-US-D index, 0.915 for the EPU-US-M index, 0.801 for the EPU-US-News-M index, and 0.924 for the EPU-CN-M index. It shows that all the EPU indices exhibit very strong persistence, which can be seen in Fig.~\ref{Fig:EPUNetStat:RawData} through eyeballing. It means that the economic policy uncertainty is persistent in each country.


\subsection{Degree distributions}

The degree $k$ of a node in a network is the number of nodes connected to it. As shown in Table~\ref{TB:EPUNetStat:TopoStat}, the minimum degree $k_{\min}$ of each EPU network is 2, indicating that the endpoints of the time series are visible to at least one additional point other than their neighbor nodes. For the networks converted from the monthly EPU series, the maximum degrees $k_{\max}$ is close to 56 and the average degree $\langle{k}\rangle$ is close to 8. In contrast, for the daily data, the maximum degree is much larger, but the average degree is smaller. 

\begin{table}[h]
\centering
\caption{Topological statistics of the four EPU networks. $k$ is the degree, $\gamma$ is tail exponent of the degree distribution, $C$ is the clustering coefficient, and $r$ is degree-degree correlation coefficient.}
\label{TB:EPUNetStat:TopoStat}
\smallskip
\begin{tabular}{ccd{3.0}cccccccc}
\hline\hline
 Network      &  $\langle{k}\rangle$ & \multicolumn{1}{c}{$k_{i,\max}$}  & $k_{i,\min}$ & $\gamma$ &  $C$  &  $C_{i,\max}$ & $C_{i,\min}$ & $r$ \\ \hline
 EPU-US-M       &  7.86  &   56  &   2   & 1.99 &   0.76   &  1   & 0.11    &  0.08    \\ 
 EPU-US-News-M  &  7.97  &   56  &   2   & 2.13 &   0.77   &  1   & 0.11    &  0.04     \\ 
 EPU-CN-M       &  8.06  &   59  &   2   & 1.83 &   0.76   &  1   & 0.13    &  0.07    \\ 
 EPU-US-D       &  6.99  &  158  &   2   & 2.78 &   0.77   &  1   & 0.05    &  0.12   \\ 
 \hline\hline
\end{tabular}
\end{table}

\begin{figure}[h!tbp]
\centering
\includegraphics[width=0.45\linewidth, height=0.32\linewidth]{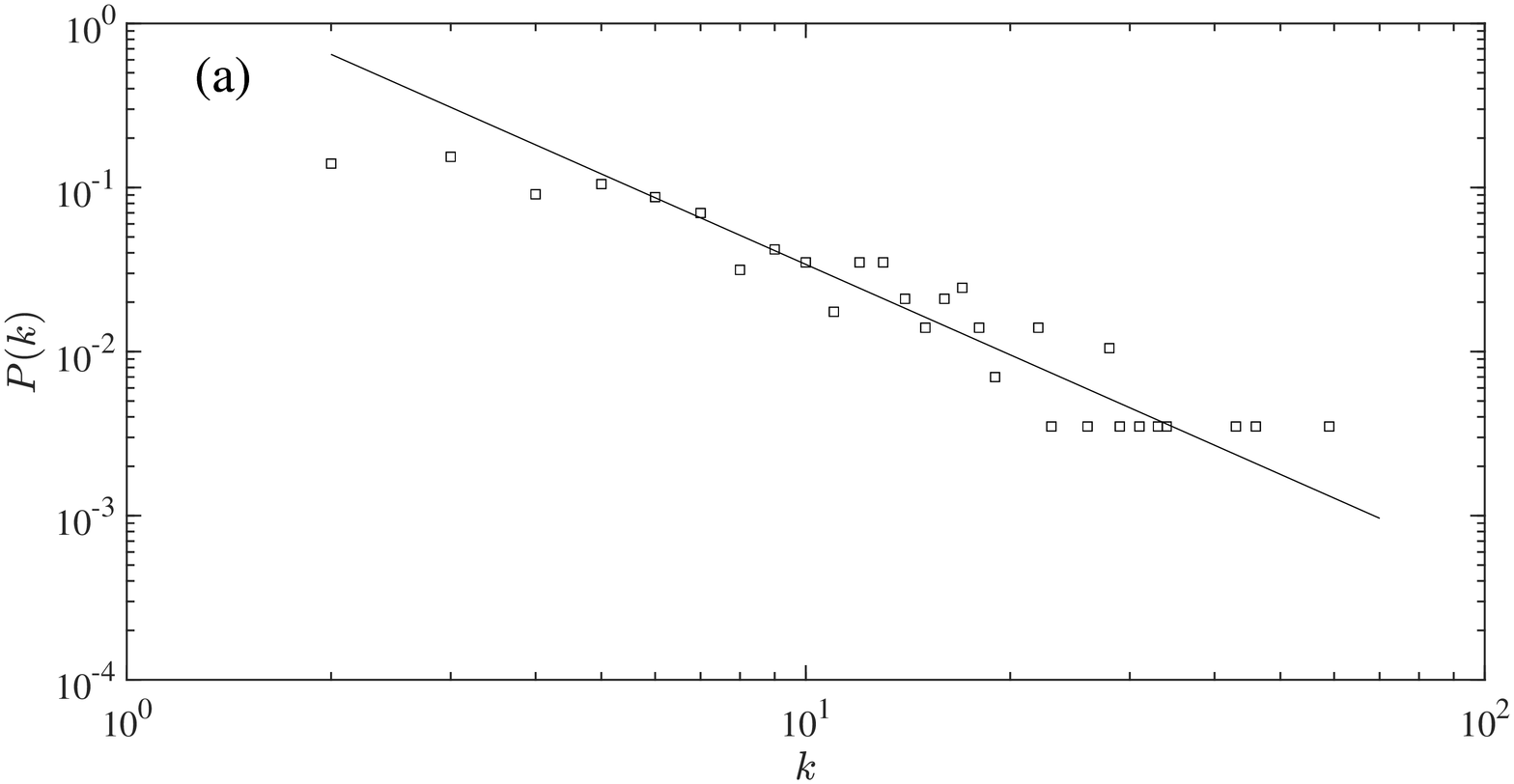}
\includegraphics[width=0.45\linewidth, height=0.32\linewidth]{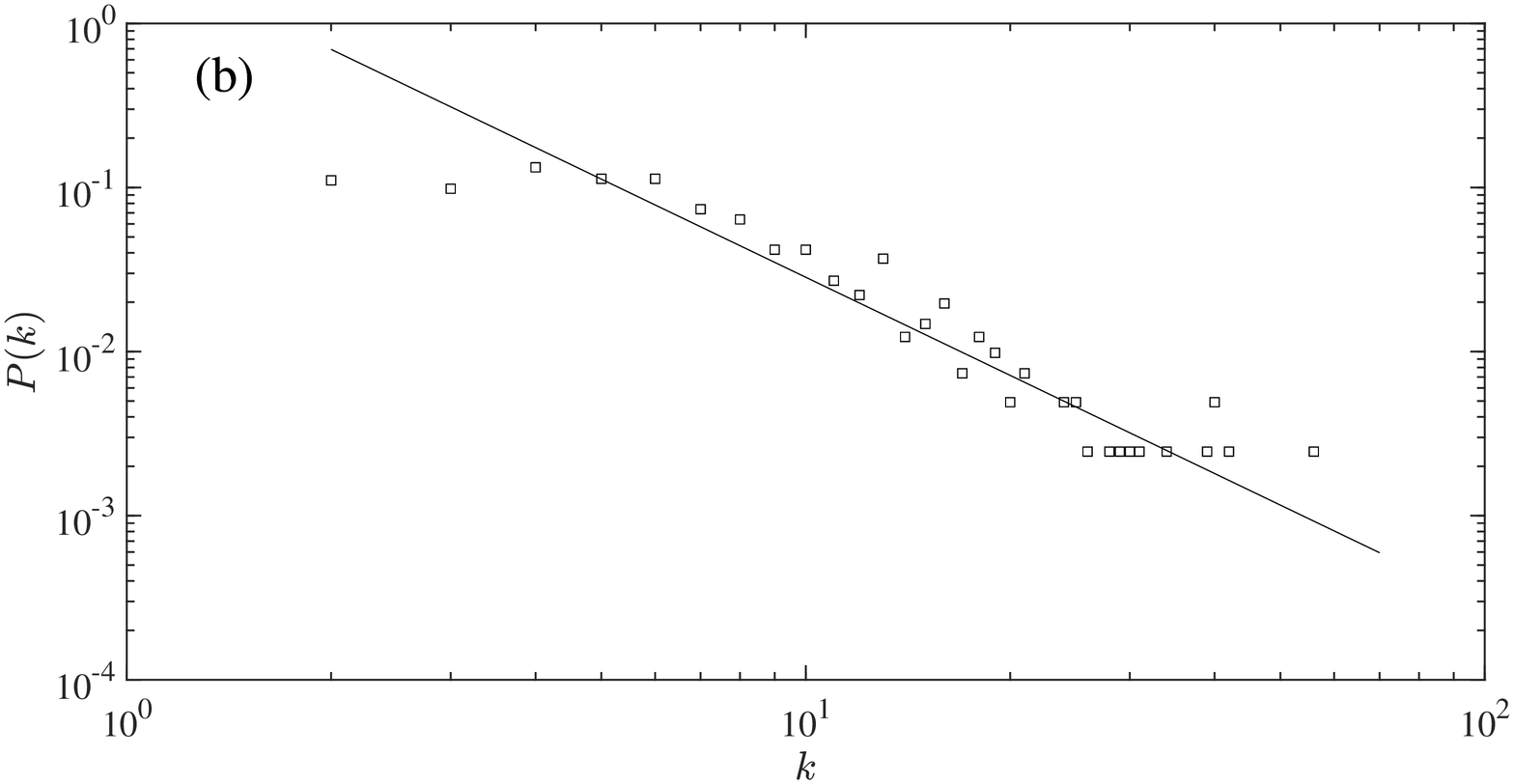}
\includegraphics[width=0.45\linewidth, height=0.32\linewidth]{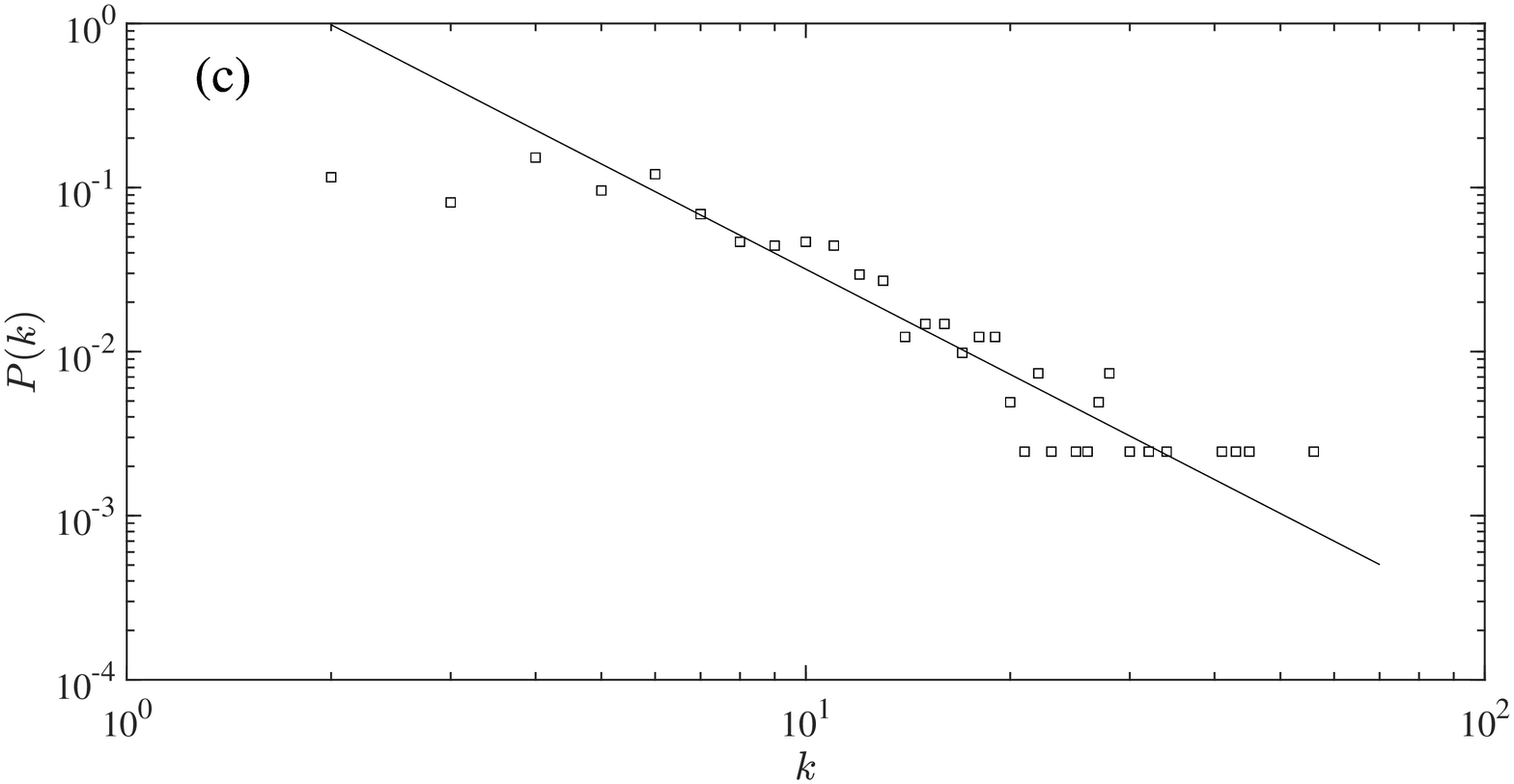}
\includegraphics[width=0.45\linewidth, height=0.32\linewidth]{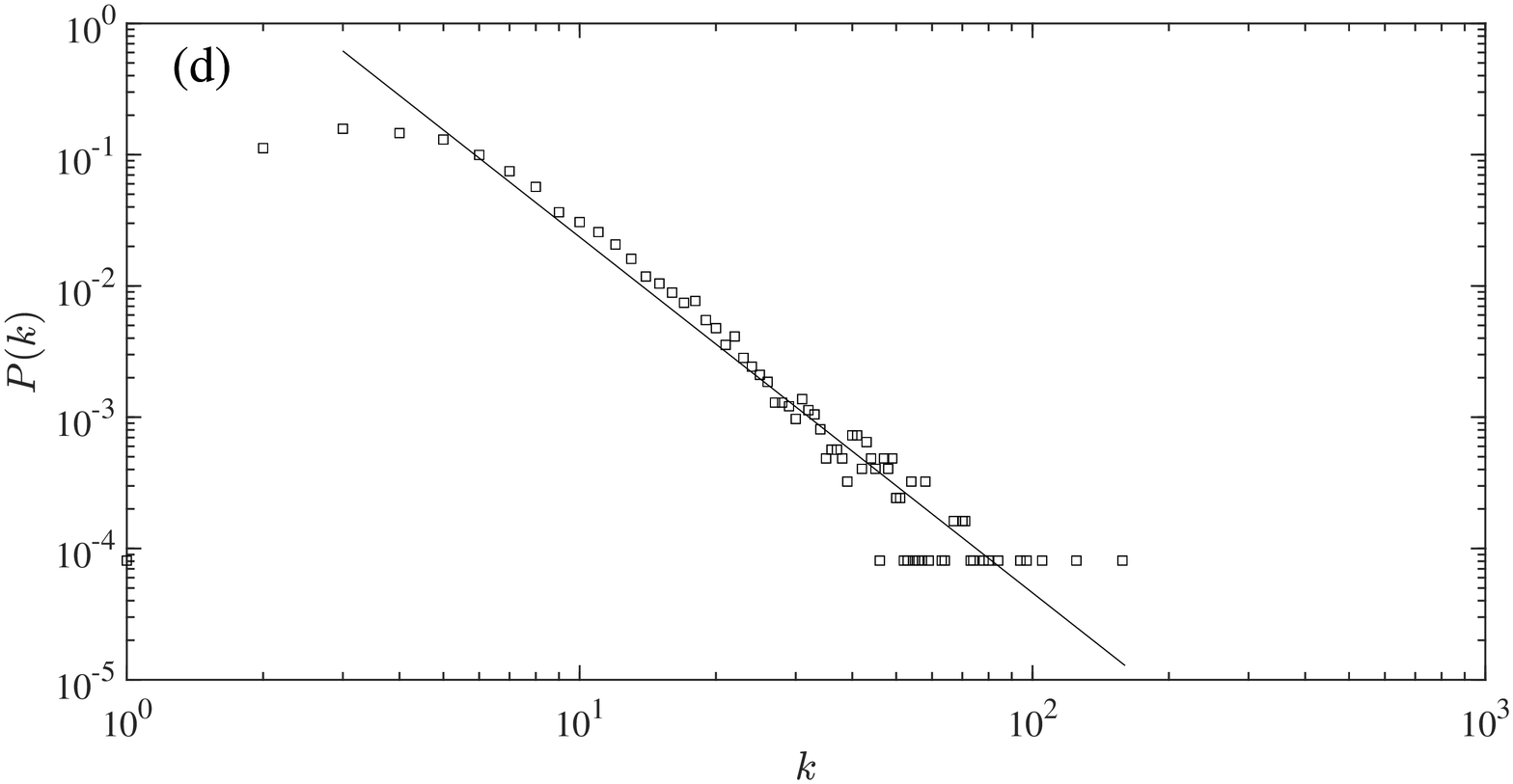}
\caption{The degree distributions of the networks converted from the four EPU time series: (a) EPU-CN-M, (b) EPU-US-M, (c) EPU-US-News-M, and (d) EPU-US-D.}
\label{Fig:EPUNetStat:PDF:Degree}
\end{figure}

We estimate the empirical degree distributions of the four EPU networks, as illustrated in Fig.~\ref{Fig:EPUNetStat:PDF:Degree}. It shows that all the distributions have power-law tails:
\begin{equation}
p(k)\sim k^{-\gamma},
\end{equation}
where $\gamma$ is the power-law tail exponent. The tail exponents are presented in Table~\ref{TB:EPUNetStat:TopoStat}. The goodness-of-fit values of the four linear regressions are 0.8703 in Fig.~\ref{Fig:EPUNetStat:PDF:Degree}(a), 0.9184 in Fig.~\ref{Fig:EPUNetStat:PDF:Degree}(b), 0.9057 in Fig.~\ref{Fig:EPUNetStat:PDF:Degree}(c), 0.9539 in Fig.~\ref{Fig:EPUNetStat:PDF:Degree}(d) respectively. Therefore, the EPU networks are scale-free.

\subsection{Clustering coefficient}

The clustering coefficient measures the cliquishness of a node in a network \cite{Watts-Strogatz-1998-Nature}. For any node $V$ in a network, its clustering coefficient is defined as follows:
\begin{equation}
 C_i=\frac{2E_i}{k_i(k_i-1)},
\end{equation}
where $k_i$ is the degree of node $i$ and $E_i$ is the number of edges between $i$'s neighbor points. We obtain all the $C_i$ values. According to Table~\ref{TB:EPUNetStat:TopoStat}, the maximum clustering coefficient is 1, which is somehow trivial, showing that there are local cliques. The minimum clustering coefficients are close and small for the three networks mapped from the monthly indices, and that from the daily indices is smaller.

The clustering coefficient of the network is defined as the average clustering coefficient of all nodes:
\begin{equation}
 C=\langle{C_i}\rangle_i=\frac{1}{N}\sum_{i=1}^NC_i,
\end{equation}
which measures the probability of triadic closure in the network. If $C=1$, the network is a complete graph, where any two nodes of the network are connected to each other. According to Table~\ref{TB:EPUNetStat:TopoStat}, the average cluster clustering coefficient is large (about 0.76) for all the EPU networks. This is consistent with the strongly persistent feature of the indices. It also shows that the edges of the EPU networks are dense.


\subsection{Assortative mixing}

Mixing pattern is an important topological statistic of complex networks, where the assortativity refers to the fact that nodes in a network are more likely to be connected with nodes of comparable degrees \cite{Newman-2002-PRL}. In assortative networks, nodes with high degrees are more likely to connect to other nodes also with high degrees, while nodes with low degrees are more likely connected to low-degree nodes. Similarly, nodes connected to nodes with lower degree may also have lower degree. The mixing pattern is generally quantified by the Pearson correlation coefficient between node degrees:
\begin{equation}
r=\frac{M^{-1}\sum_ij_ik_i-\left[M^{-1}\sum_i\frac{1}{2}\left(j_i+k_i\right) \right]^2}
       {M^{-1}\sum_i\frac{1}{2}\left(j_i^2+k_i^2\right)-\left[M^{-1}\sum_i\frac{1}{2}\left(j_i+k_i\right) \right]^2}
\end{equation}
where $M$ is the number of all edges in the network, $j_i$ and $k_i$ are the degrees of the endpoint at the $i$th edge. If $r$ is zero, the network has no assortative mixing. If $r$ is positive, the network exhibits assortative mixing. If $r$ is negative, the network shows disassortative mixing.


Table~\ref{TB:EPUNetStat:TopoStat} lists the assortativity coefficients of the four economic policy uncertainty networks. All the assortative coefficients are positive but small, which implies that the four networks exhibit weak assortative mixing patterns. The assortativity coefficient of EPU-US-D is larger than the other three ones. In the network corresponding to higher-frequency economic policy uncertainty index, there exists relatively stronger assortativity. In terms of economic policy uncertainty, the networks from monthly data have only weak assortativity. In networks from the visibility algorithm, the larger the value of EPU index is, the higher degree of the corresponding node will be to a large extend. A considerable degree of economic policy uncertainty in the economic environment will not disappear quickly in the short term. Hence, there will be a period of high economic policy uncertainty which explains why the day-based assortativity coefficient is higher than the month-based assortativity coefficient.

\subsection{Small-world network}

A network with the small-world property usually has two main characteristics \cite{Watts-Strogatz-1998-Nature}. Firstly, the average shortest path length $L\left(N\right)$ of a network fulfills:
\begin{equation}
  L(N)\sim \ln N
\end{equation}
where $N$ is the number of nodes in the network, and $L\left(N\right)$ is calculated as follows:
\begin{equation}
  L\left(N\right)=\frac{2}{N(N-1)}\sum{d\left(i,j\right)}
\end{equation}
where $d\left(i,j\right)$ is the shortest path length (or distance) between any two nodes $i$ and $j$. Secondly, the clustering coefficient of the network should be large.

\begin{figure}[htbp]
\centering
\includegraphics[width=0.5\linewidth,height=0.3\linewidth]{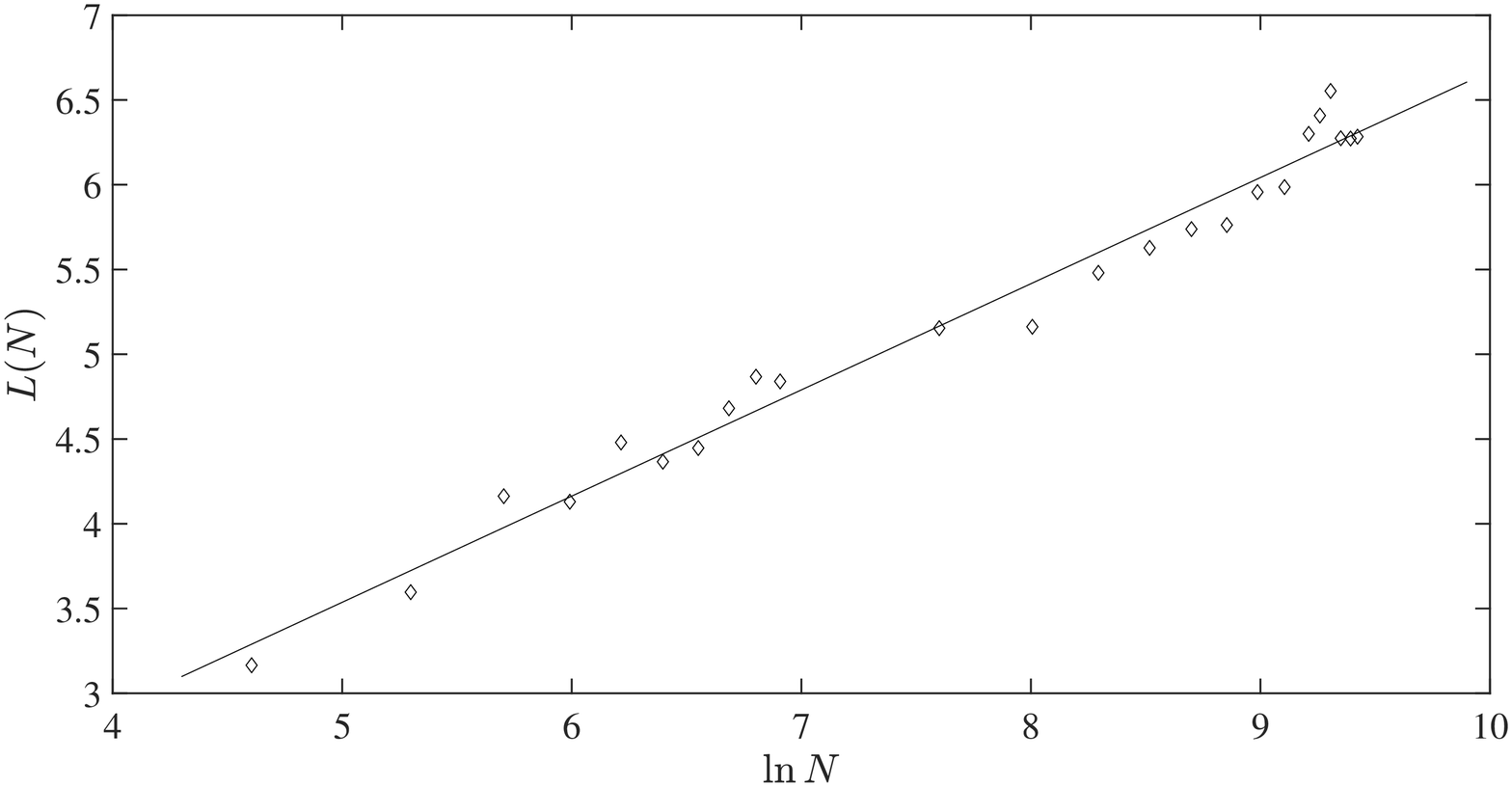}
\caption{Dependence of the average shortest path length on the total number of nodes for the EPU network converted from the daily US data.}
\label{Fig:EPUNetStat:ASD}
\end{figure}

Fig.~\ref{Fig:EPUNetStat:ASD} shows the dependence of the average shortest path length of the network on the total number of nodes for the EPU network converted from the daily US data. We do not analyze the networks from the monthly data because their sizes are not large enough. A linear regression gives that $L\left(N\right)=0.626\ln N+0.405$. On the other hand, the clustering coefficient the network is 0.77, which is a large value. Therefore, the EPU network of the daily US index is a small-world network.

\section{Conclusions}
\label{S1:Conclude}		

This paper utilizes the EPU indices as a proxy variable of economic policy uncertainty. The data shows that China has much higher and more volatile economic policy uncertainty than the United States. The Hurst exponents of the EPU index time series in the United States and China have been calculated, which are found to be strongly long-range correlated. It suggests that the economic policy uncertainty is persistent in different economies.

The EPU indices are mapped into complex networks by the visibility graph algorithm. We studied the topological properties of these networks. By investigating the degree distributions, the clustering coefficients, the mixing patterns, and the shortest path lengths, we unveiled that the EPU networks are scale-free, densely connected and weakly assortative. The EPU network constructed from the daily US index also exhibit evident small-world features.

Our research highlights the possibility to study the EPU from the view angle of complex networks. For single time series, we can convert them into networks using different methods. The visibility graph approach used in this work is only one of them. Different mapping methods result in different networks. It is thus possible to unveil different structural properties that may correspond to different temporal features of the time series. We can also consider cross-sectional methods such as the random matrix theory, which aims at understanding the global and regional behavior of economic policy uncertainty. These issues can be investigated in the future research.



\section*{Acknowledgements}

This work was supported by the National Natural Science Foundation of China (Grants No. 71532009, U1811462 and No. 71790594), the Fundamental Research Funds for the Central Universities, and Tianjin Development Program for Innovation and Entrepreneurship.

\bibliography{Bibliography}

\end{document}